\documentclass[aps,pre,showpacs,superscriptaddress,final,floatfix]{revtex4-2}
\usepackage{setspace}
\usepackage{amssymb}
\usepackage{amsmath}
\usepackage{color}
\usepackage{textcomp}
\usepackage[utf8]{inputenc}

\usepackage{graphicx}
\graphicspath {{./Figures/}}
\usepackage{textcomp}
\usepackage[
colorlinks=true, 
pdfstartview=FitV, 
linkcolor=magenta, 
citecolor=blue, 
urlcolor=black, 
bookmarks=true,
bookmarksnumbered=true,
pdftitle={},
pdfauthor={}
]{hyperref}

\usepackage{amsmath}

\begin{document}
\linespread{1.3}\selectfont{}

	\title{Quasiperiodicity in the $\alpha-$Fermi-Pasta-Ulam-Tsingou problem revisited:  an approach using ideas from wave turbulence }
	
	\author{Santhosh Ganapa}
	\email{sunny.ganapa@kuleuven.be}
	\email{santhosh.eeebitspilani@gmail.com}

	\affiliation{Institute for Theoretical Physics, KU Leuven -- 3001, Belgium}
	\affiliation{International Centre for Theoretical Sciences $-$ Tata Institute of Fundamental Research, Bengaluru -- 560089, India}

	\begin{abstract} The Fermi-Pasta-Ulam-Tsingou (FPUT) problem addresses fundamental questions in statistical physics, and attempts to understand the origin of recurrences in the system have led to many great advances in nonlinear dynamics and mathematical physics. In this work we revisit the problem and study quasiperiodic recurrences in the weakly nonlinear $\alpha-$FPUT system in more detail. 
	We aim to reconstruct the quasiperiodic behaviour observed in the original paper from the canonical transformation used to remove the three wave interactions, which is necessary before applying the wave turbulence formalism. We expect the construction to match the observed quasiperiodicity 
	if we are in the weakly nonlinear regime. 
	Surprisingly, in our work we show that this is not always the case and in particular, the recurrences observed in the original paper cannot be constructed by our method. 
	We attribute this disagreement to the presence of small denominators in the canonical transformation used to remove the three wave interactions before arriving at the starting point of wave turbulence. We also show that these small denominators are present even in the weakly nonlinear regime, and they become more significant as the system size is increased. 
	We also discuss our results in the context of the problem of equilibration in the $\alpha-$FPUT system, and point out some mathematical challenges when the wave turbulence formalism is applied to explain thermalization in the $\alpha-$FPUT problem. We argue that certain aspects of the $\alpha-$FPUT system such as presence of the stochasticity threshold, thermalization in the thermodynamic limit and the cause of quasiperiodicity are not clear, and that they require further mathematical and numerical studies.
	
\end{abstract}

\date{13 Apr 2023}
\maketitle

\section{Introduction}
\label{sec:Intro}
The Fermi–Pasta–Ulam–Tsingou (FPUT) problem is a classic study with a long history. It was originally studied in order to understand the cause of thermalization in a macroscopic system. The expectation when the problem was first posed \cite{Fermi1955,dauxois2008fermi} was that nonlinearity should be enough for a system to thermalize. But instead, quasiperiodic behaviour was observed with near perfect recurrences to the initial state. At the time the authors of the paper had remarked: ``Let us say here that the results of our computations show features which were, from the beginning, surprising to us''. But this ``surprise'' actually lead to a number of studies in the following years aimed at understanding their results, which led to significant developments in statistical physics, nonlinear dynamics and mathematical physics. As far as the question of if and when the FPUT chain thermalizes is concerned, there have been many studies done to investigate this. Some of the methods include the closeness to integrability and soliton solutions \cite{Zabusky1965}, stochasticity threshold \cite{Chirikov1960,Izrailev1966}, Lyapunov exponent studies \cite{Casetti1996,Benettin2018, Liu2021}, breather solutions \cite{Marin1996, Flach2005, Flach2006,Danieli2017} and local equilibration studies \cite{Ganapa2020}. More recently the formalism of wave turbulence \cite{Zakharov1992,Nazarenko2011} was applied to the FPUT chain \cite{Onorato2015,Lvov2018,Pistone2019, Bustamante2019} in order to explain the dependence of equilibration time on the nonlinearity parameter.

In this work, we study quasiperiodicity in the $\alpha-$FPUT system (FPUT system with cubic nonlinearity) in more detail. 
Our goal is to understand the cause of quasiperiodic recurrences. Previous studies \cite{Onorato2015} attribute the quasiperiodic behaviour (such as the one observed in the original paper \cite{Fermi1955}) to the lack of three wave resonant interactions. 
The validity of this statement is important in order for the wave turbulence description to be applicable for this system. The implication of this statement is that the three wave interactions can be removed by a suitable canonical transformation. Removing the three wave interactions is necessary before we start applying the methods of wave turbulence. This process is described in \cite{Onorato2015}, and we will discuss this briefly. In this work, we aim to test if the quasiperiodic behaviour can indeed be attributed to the  lack of three wave resonances. We do this by checking if the canonical transformation used to remove the three wave interactions has information about the quasiperiodic behaviour. We use the canonical transformation to construct the evolution of the $\alpha-$FPUT system at short times. We also use the formalism of wave turbulence to compute the frequencies of the $\alpha-$FPUT chain in terms of perturbative corrections to the harmonic chain, which is necessary in order to construct the evolution even more accurately. We then compare this evolution with that of molecular dynamics simulations. We obtain a good agreement between the two for smaller system sizes and nonlinearities, which suggests that the information about quasiperiodicity is present in the canonical transformation. However, the agreement between them gets poorer as the system size increases. We then explain this lack of agreement in terms of small denominators present in the canonical transformation, which lead to a regime where the lack of three wave resonances is not sufficient to explain the quasiperiodic behaviour. We argue and provide numerical evidence that there is a divergence in the canonical transformation leading to a breakdown of the perturbation theory, and discuss the onset of strong nonlinearity. We point out that we run into difficulties in this regime even before we start applying the wave turbulence formalism, and argue that we need to understand other physical processes which may be responsible for quasiperiodicity observed in the original paper \cite{Fermi1955}.


This paper is organized as follows: in Sec.~\ref{sec:Model}, we introduce the $\alpha-$FPUT problem and describe it in terms of variables that are easier to work in the context of wave turbulence. 
Our main work is first briefed in Sec.~\ref{sec:Method} and then described in Sec.~\ref{sec:Solution}, where we construct an analytic expression for quasiperiodicity in the $\alpha-$FPUT system by using the canonical transformation used to remove the three wave interactions (and also other non resonant interactions). This canonical transformation is basically a perturbation series in terms of the nonlinearity parameter $\epsilon$, which we define later. We use terms up to first order in our study. In Sec.~\ref{sec:Numerics} we numerically compare our expression with the dynamics of the $\alpha-$FPUT system obtained by evolving the Hamilton's equations of motion. We go on to discuss the anomalies in terms of a potential breakdown of the perturbation series. 
We finally discuss our results in the context of thermalization in the $\alpha-$FPUT system and conclude our discussion in Sec.~\ref{sec:Conclusion}. The calculation used to derive the frequencies of the $\alpha-$FPUT chain is presented in Appendix.~\ref{app:freq}.

\section{Model and Review of Earlier Work} 
\label{sec:Model}
\subsection{The FPUT problem}
The $\alpha-$FPUT system of $N$ particles, each of mass $m$ is described by the Hamiltonian as a function of the position $q_i$ and momentum $p_i$ of each particle as:
\begin{equation}\label{eq:hamFPU}
H(\boldsymbol{p},\boldsymbol{q})=\sum_{i=0}^{N-1}\left[\frac{p_i^2}{2m}+\frac{\mu (q_{i+1}-q_{i})^2}{2}+\frac{\alpha(q_{i+1}-q_{i})^3}{3}\right]~,
\end{equation}
where we have considered periodic boundary conditions with $q_N \equiv q_0$ and $q_{-1} \equiv q_{N-1}$. If $\alpha=0$, the system reduces to the harmonic chain with $\mu$ being the spring constant. From now on we take $m = \mu = 1$ unless otherwise mentioned. For the harmonic chain, one can make a canonical transformation from $p,q$ variables to the normal mode variables $P,Q$ defined by:
\begin{equation}\label{eq:nm}
Q_k =\frac{1}{N} \sum_{j=0}^{N-1}q_je^{-i2\pi kj/N},~~ P_k =\frac{1}{N} \sum_{j=0}^{N-1}p_je^{-i2\pi kj/N}.  
\end{equation}
The  Hamiltonian then becomes
\begin{align}
H=N\sum_{k=0}^{N-1} E_k,~~{\rm where}~ E_k=\frac{|P_k|^2}{2}+\frac{\omega_k^2 |Q_{k}|^2}{2}~ \label{HNM}
\end{align}
is the energy of each mode and $\omega_k = 2 \sin(k\pi/N), k = 0,1,2,...,N-1$ is the normal mode frequency of the $k^{th}$ mode. The zero mode corresponding to $k=0$ does not participate in the dynamics, and we only deal with initial conditions that have $E_0 = 0$ (zero initial momentum). From Hamilton's equations of motion in these canonical coordinates one can observe that these normal modes decouple. 
So, energy exchange does not take place between different normal modes. Hence, most of the phase space is not accessible to a harmonic chain and the system does not thermalize. We would like the normal modes to somehow interact with each other in order to expect thermalization. For that we add a cubic interaction term to the Hamiltonian ($\alpha \neq 0$). 
We then get the $\alpha-$FPUT system described by the Hamiltonian Eq.~\eqref{eq:hamFPU}. The origin of this Hamiltonian can be understood as arising from a spring that has restoring force $F$ dependent on the displacement $x$ of the particle connected to it as $F = - \mu x -\alpha x^2$. Thus, $\alpha$ is the origin of nonlinearity in the system. Note that the cubic potential of the $\alpha$-FPUT system implies that the system stays bounded only if the total energy  is sufficiently small and the precise condition is $E< \mu^3/(6 \alpha^2)$, corresponding to all energy contributing to the potential energy of a single particle \cite{Ganapa2020}. In practice this is highly improbable and one can work with energies slightly higher than this bound. This restriction is however not needed in the $\beta$-FPUT system, which has $F = - \mu x -\beta x^3$, where $\beta$ is strength of the nonlinearity. The $\alpha$-FPUT system is studied inspite of this restriction because we can observe higher order effects more easily than the $\beta$ system (one of them is the fact that the first order correction to the harmonic frequency is zero for the $\alpha$-FPUT system (as we will see later in Appendix.~\ref{app:freq}), while it is non-zero for the $\beta$-FPUT system, which has been computed in \cite{Lvov2018}).

For sufficiently small nonlinearity, the energy contribution from nonlinear part of the interaction potential is small and it is a good approximation to assume that the total energy can still be approximated as a sum of energies of the independent harmonic oscillators, i.e, the total energy $E \approx \sum E_k$. In that case, one check of equipartition would be to see if all the $E_k(t)$ converge, at long times, to the same value $e=E/(N-1)$ (perhaps with small fluctuations). This was the approach in \cite{Fermi1955} (where however the fixed boundary condition case was studied). There, energy was initially given to the first normal mode and the time evolution was studied numerically. Contrary to the expectations, the long-time dynamics appeared to be almost periodic, with near perfect returns to the initial condition. The original paper also considered the $\beta$-FPUT system and broken linearity. In all these cases they observed quasiperiodicity instead of thermalization. 

Since the original paper, many different approaches were developed in order to explain the absence of thermalization in the FPUT system, which is sometimes referred to as the FPUT paradox. The FPUT problem has a vast literature and we only refer to some of the review articles \cite{Ford1992, Weissert1997, Berman2005,Gallavotti2008,Benettin2013}. Zabusky and Kruskal have explained this paradox by relating this to the closeness of FPUT system to a completely integrable model \cite{Zabusky1965} in the continuum limit known as the Korteweg–de Vries (KdV) equation, which has soliton solutions. Another approach investigated the role of breather solutions \cite{Marin1996,Flach2005,Flach2006,Christodoulidi2010,Danieli2017} (time-periodic and space-localized solutions) in delaying and even preventing thermal behaviour. 
In a certain sense the idea is similar to the one relating the presence of solitons in the KdV system to the absence of equilibration in the FPUT system $-$ the difference being that breathers are stable solutions of the discrete system, while the KdV is a continuum approximation. 
A different approach in resolving the FPUT paradox was developed by Chirikov and Izrailev \cite{Chirikov1960, Izrailev1966} on the basis of the criterion of overlapping of resonances, leading to the stochasticity threshold.  According to these studies, the initial conditions used by FPUT in their numerical simulations were chosen below the stochasticity threshold, just in the region corresponding to stable quasi-periodic motion. For the harmonic chain, for small values of $k(k \ll N$), the separation between successive levels scales as $\Delta_\omega \sim 1/N$, while for $N-k \ll N$, $\Delta_\omega \sim 1/N^2$. Hence the stochasticity threshold is larger for low frequency modes. Since the FPUT study had initial conditions with the lowest mode excited and the energy density was small, it is plausible that they were below the threshold. Above this threshold, the FPUT model was shown to behave in accordance with the original expectations of FPUT, revealing strong statistical properties such as energy equipartition among the linear modes. 
This idea has also been studied numerically \cite{Livi1985,Deluca1995,Casetti1996}, where attempts were  made to relate thermalization in the system to the untrapping of the system from its regular region and escape to the chaotic component of its phase space. 
A recent work done on the $\alpha-$FPUT system \cite{Ganapa2020} studied local equilibration in the system, and investigated the role of initial conditions, choice of observables, choice of the averaging protocol and the role of chaos in thermalization. The FPUT paradox has also been studied by using the methods of wave turbulence \cite{Onorato2015,Lvov2018,Pistone2019, Bustamante2019}, which we will discuss after describing the evolution of $\alpha-$FPUT system in a language that is more appropriate to the formalism of wave turbulence.
\subsection{$\alpha-$FPUT system in the language of wave turbulence}

Wave turbulence \cite{Zakharov1992,Nazarenko2011} refers to the statistical theory of weakly nonlinear dispersive waves. 
Wave turbulence arises in a wide variety of contexts. For example, wave turbulence arises in surface waves on water (both gravity and capillary) \cite{Dyachenko1994,Dyachenko1995,Brazhnikov2002,Lukaschuk2009,Cobelli2011,Falcon2022}, nonlinear optical systems \cite{Dyachenko1992,Bortolozzo2009}, sound waves in oceanic waveguides \cite{Gurbatov2005}, shock waves in the solar atmosphere and their coupling to the Earth’s magnetosphere \cite{Ryutova2003}, and magnetic turbulence \cite{David2022} in interstellar gases \cite{Bisnovatyi1995}. The formalism of wave turbulence has also been used to explain anomalous conduction in one-dimensional particle lattices \cite{Devita2022}. Recently, the problem of equilibration in the FPUT system has been approached  by using ideas from wave turbulence \cite{Onorato2015,Lvov2018,Pistone2019, Bustamante2019}. The idea of the approach is to  connect the equilibration issue to the presence of higher order resonances between dressed normal modes that appear under repeated canonical transformations. 
 Let us now study the $\alpha-$FPUT system in the language of wave turbulence. We mostly repeat the calculations given in \cite{Onorato2015} here. First let us describe the $\alpha-$FPUT system in terms of the normal mode coordinates $a_k$:
\begin{equation}
a_k = \frac{1}{\sqrt{2\omega_k}}(P_k-i\omega_kQ_k).
\end{equation} 
Writing in terms of the dimensionless variables,
$$ a_k^\prime = \frac{(\mu/m)^{1/4}}{\sqrt{\sum_k \omega_k \mid a_k(t=0)\mid^2 }}a_k,\ t^\prime = \sqrt{\frac{\mu}{m}}t,\ \omega_k^\prime=\sqrt{\frac{m}{\mu}}\omega_k,$$
the Hamiltonian Eq.~\eqref{eq:hamFPU} can be written in terms of the canonical variables $\{ia_k,a_k^\star\}$ (after removing primes for brevity) as:
\begin{equation}
\begin{split}
\frac{H}{N}=\sum_{k=1}^{N-1}\omega_k a_k^\star a_k +\epsilon\sum_{k_1,k_2,k_3}V_{k_1,k_2,k_3}[\frac{1}{3}(a_{k_1}a_{k_2}a_{k_3}+a_{k_1}^\star a_{k_2}^\star a_{k_3}^\star)\delta_{k_1,-k_2-k_3} \\ +(a_{k_1}^\star a_{k_2}a_{k_3}+a_{k_1} a_{k_2}^\star a_{k_3}^\star)\delta_{k_1,k_2+k_3}]~.
\end{split}
\end{equation}
The matrix $V_{k_1,k_2,k_3}$ weights the transfer of energy between wave numbers $k_1$, $k_2$ and $k_3$ and is defined as:
\begin{equation}\label{eq:transfer}
	V_{k_1,k_2,k_3} = -\frac{1}{2\sqrt{2}}\frac{\sqrt{\omega_{k_1}\omega_{k_2}\omega_{k_3}}}{sign(sin(\frac{\pi k_1}{N})sin(\frac{\pi k_2}{N})sin(\frac{\pi k_3}{N}))}~.
\end{equation}
The dimensionless parameter $\epsilon$ is given by:
\begin{equation}\label{eq:nonlinear}
	\epsilon = \frac{\alpha}{m}(\frac{\mu}{m})^{1/4}\sqrt{\sum_k \omega_k \mid a_k(t=0)\mid^2 } ~.
\end{equation}
This leads to the equations of motion:
\begin{equation}\label{eq:evol}
i\frac{\partial a_{k_1}}{\partial t} = \frac{1}{N}\frac{\partial H}{\partial a_{k_1}^*}=\omega_{k_1} a_{k_1} + \epsilon \sum_{k_2,k_3} V_{k_1,k_2,k_3} (a_{k_2}a_{k_3}\delta_{k_1,k_2+k_3}+2a_{k_2}^\star a_{k_3}\delta_{k_1,k_3-k_2}+a_{k_2}^\star a_{k_3}^\star\delta_{k_1,-k_2-k_3})~.
\end{equation}
The delta function is 1 also if the argument differs mod N. 
The variables $k_2$ and $k_3$ run from $-N/2+1$ to $N/2$. 
Eq.~\eqref{eq:evol} describes the evolution of the $\alpha-$FPUT system that is more apt to the formalism of wave turbulence. 
We are now interested in the weak nonlinearity limit, i.e., $\epsilon \ll 1$.

One can observe from Eq.~\eqref{eq:evol} that the nonlinear behaviour of the $\alpha$-FPUT system is due to the three-wave interactions between $a_{k_1}$, $a_{k_2}$ and $a_{k_3}$, with the strength of coupling determined by $\epsilon$.  First we find out if these three waves interactions are resonant. The way to check for resonances is to do a canonical transformation and then try to remove the three wave interactions. If the canonical transformation does not have a zero denominator then we say that the three wave interactions are not resonant. Otherwise we conclude that there are resonances, and that these interactions cannot be removed by any canonical transformation. This canonical transformation is of the form:

\begin{equation}\label{eq:ct}
a_{k_1}=b_{k_1}+\epsilon\sum_{k_2,k_3}(A_{k_1,k_2,k_3}^{(1)}b_{k_2}b_{k_3}\delta_{k_1,k_2+k_3}+A_{k_1,k_2,k_3}^{(2)}b_{k_2}^\star b_{k_3}\delta_{k_1,k_3-k_2}+A_{k_1,k_2,k_3}^{(3)}b_{k_2}^\star b_{k_3}^\star\delta_{k_1,-k_2-k_3})+O(\epsilon^2)~.
\end{equation}
Substituting this in Eq.~\eqref{eq:evol} and equating the coefficient of $\epsilon$ on both the sides, we get:
$$ A_{k_1,k_2,k_3}^{(1)} = V_{k_1,k_2,k_3}/(\omega_{k_3}+\omega_{k_2}-\omega_{k_1}),$$
$$ A_{k_1,k_2,k_3}^{(2)} =2V_{k_1,k_2,k_3}/(\omega_{k_3}-\omega_{k_2}-\omega_{k_1}),$$
$$ A_{k_1,k_2,k_3}^{(3)} =V_{k_1,k_2,k_3}/(-\omega_{k_3}-\omega_{k_2}-\omega_{k_1}).$$
It is easy to verify using trigonometric identities that for the frequencies $\{\omega_k=2\sin(k\pi/N)\} $ of the harmonic chain 
the denominators in the above transformation $\{\omega_{k_3}+\omega_{k_2}-\omega_{k_1}$, $\omega_{k_3}-\omega_{k_2}-\omega_{k_1}$, $\omega_{k_3}+\omega_{k_2}+\omega_{k_1}\}$ are never zero. Therefore, the canonical transformation Eq.~\eqref{eq:ct} from $a$ to $b$ variables is well defined, the three wave interactions are not resonant and can be removed by the canonical transformation Eq.~\eqref{eq:ct}. This lack of three wave resonances has been linked \cite{Onorato2015} to the quasiperiodic behaviour at short time scales observed in the original paper \cite{Fermi1955}. We will come back to this point later in our work. We also point out that the denominators can get very small, although they are never zero. We will postpone further discussion on this until Sec.~\ref{sec:Numerics}. Now substituting the transformation Eq.~\eqref{eq:ct} in the equations of motion Eq.~\eqref{eq:evol}, we get:
\begin{equation}\label{eq:evol2}
i\frac{\partial b_{k_1}}{\partial t} = \omega_{k_1} b_{k_1} + \epsilon^2 \sum_{k_2,k_3,k_4} T_{k_1,k_2,k_3,k_4} b_{k_2}^\star b_{k_3}b_{k_4}\delta_{k_1+k_2,k_3+k_4} +O(\epsilon^3)~.
\end{equation}
The above equation also has terms including the Kronecker deltas $\delta_{k_1,k_3+k_4+k_2}$, $\delta_{k_1,-k_3-k_4-k_2}$ and $\delta_{k_1,k_4-k_3-k_2}$, which are of order $\epsilon^2$. However, those terms are not resonant and can be removed by higher order terms in the transformation, and are of no use in this analysis. The matrix $T_{k_1,k_2,k_3,k_4}$ depends on $V_{k_1,k_2,k_3}$, and the exact expression described in \cite{Dyachenko1994}, is given in Appendix.~\ref{app:freq}.

Eq.~\eqref{eq:evol2} is the starting point of wave turbulence. While Eq.~\eqref{eq:evol2} describes the $\alpha-$FPUT system for the dispersion relation $\omega_k = 2sin(k\pi/N)$, its structure is the same for an arbitrary dispersion relation $\omega = \omega(k)$. For instance,  $\omega_k = \sqrt{gk}$, $k = 1,2,...$ is the dispersion relation of gravity waves and Eq.~\eqref{eq:evol2} in the continuum limit (sometimes known as the Zakharov equation) has been used to argue in support of the integrability of free-surface hydrodynamics in the one-dimensional case \cite{Dyachenko1994}. It has also been used to study the interaction of gravity waves propagating on the surface of an ideal fluid of infinite depth \cite{Dyachenko1995}. 

Eq.~\eqref{eq:evol2} has also been used to explain the route to thermalization \cite{Onorato2015} in the $\alpha-$FPUT system, which we briefly mention here. Since there are no three wave resonances in the $\alpha-$FPUT system, quasiperiodic behaviour is observed in the system up to times of order $\epsilon^{-4}$ after which, four wave interactions begin to dominate the dynamics. It can then be shown that four wave resonant interactions though present, are isolated from other quartets and cannot spread the energy across the spectrum. The six wave resonant interactions are interconnected, and are the lowest order interactions that lead to an effective irreversible transfer of energy. Using this analysis, the theory of wave turbulence predicted that the equilibration time scales as $\epsilon^{-8}$ for $N=16, 32, 64$, which was also verified numerically for $N=32$ \cite{Onorato2015}.  In the thermodynamic limit, this is predicted to change to $\epsilon^{-4}$ (and verified numerically \cite{Pistone2019}) because the four wave resonances are interconnected in the thermodynamic limit, which does not happen for finite system sizes.

\section{Brief overview of our work}
\label{sec:Method}
\begin{figure}[ht]
	\centering
	\includegraphics[width=0.7\textwidth]{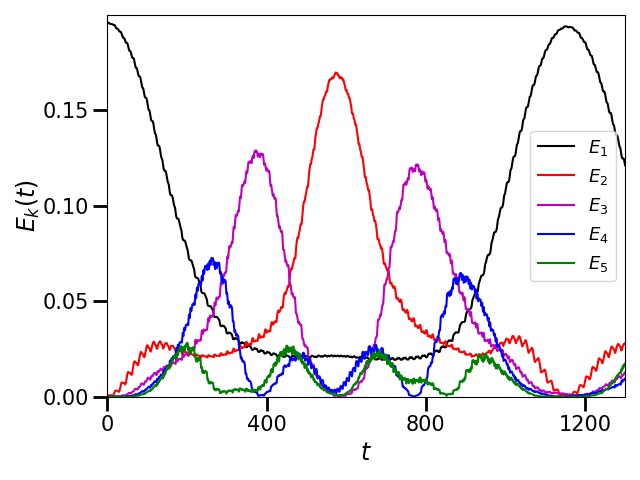}
	\caption{Quasiperiodic recurrence in the $\alpha-$FPUT problem: Plot shows the evolution of $E_k$ for different values of $k$ when only $k=1$ and $31$ are excited initially for  $N=32$ and $\epsilon = 0.05$.
	The plot closely resembles the behaviour observed by \cite{Fermi1955}, which however used fixed boundary condition. This is the well-known FPUT paradox. 
	}
	\label{FPU}
\end{figure}

The main objective of this study is to understand quasiperiodicity in the $\alpha-$FPUT system. 
Our work aims to understand the origin of quasiperiodic behaviour in the $\alpha-$FPUT system. Fig.~\ref{FPU} shows the observed quasiperiodic recurrence in the  $\alpha-$FPUT system for $N=32$ and $\epsilon = 0.05$ when the first and the last normal modes are excited initially (with amplitudes $a_1=a_{-1}=1i$). This plot closely resembles the behaviour observed in the original paper \cite{Fermi1955}, where the fixed boundary condition case was studied. 
We check if we can extract this behaviour from Eq.~\eqref{eq:evol2}. 
We rewrite Eq.~\eqref{eq:evol2} here for convenience:
$$i\frac{\partial b_{k_1}}{\partial t} = \omega_{k_1} b_{k_1} + \epsilon^2 \sum_{k_2,k_3,k_4} T_{k_1,k_2,k_3,k_4} b_{k_2}^\star b_{k_3}b_{k_4}\delta_{k_1+k_2,k_3+k_4} +O(\epsilon^3)~. $$
We now summarize our method for constructing the evolution of the $\alpha-$FPUT chain for short times (valid up to timescales of order $\epsilon^{-4}$). We are given $a_k(0)$ for all $k$ and we have to find $a_k(t)$. The basic picture that illustrates our method for finding the evolution is given below:
$$ a_k(0)\xrightarrow[\text{transformation}]{\text{canonical}} b_k(0) \xrightarrow[\text{frequency correction}]{\text{compute}}\xrightarrow[\text{evolution}]{i\frac{\partial b_k}{\partial t} \approx \Omega_k b_k}b_k(t)\xrightarrow[\text{transformation}]{\text{canonical}} a_k(t).$$ 

The idea is that $a_k$ are the normal modes of the harmonic chain and so approximating Eq.~\eqref{eq:evol} by $i\frac{\partial a_{k_1}}{\partial t} \approx \omega_{k_1} a_{k_1}$ is not a good one, for this is the evolution equation of the harmonic chain. However, approximating Eq.~\eqref{eq:evol2} by $i\frac{\partial b_{k_1}}{\partial t} \approx \omega_{k_1} b_{k_1}$ is much better because we are neglecting terms of order $\epsilon$ in the former equation and $\epsilon^2$ in the later. Its solution $b_k(t)= b_k(0)e^{-i\omega_k t}$ describes the approximate evolution of the $\alpha-$FPUT system that is valid up to times of order $\epsilon^{-4}$. While computing the evolution of $b_k$ we also have to renormalize the dispersion relation for consistency of the wave turbulence formalism \cite{Zakharov1999,Nazarenko2011,Onorato2020}. If not included in the renormalized dispersion relation, the nonlinear corrections to the frequency would lead to an unphysical secular growth in the evolution. 
So the evolution of $b_k$ is given by: $b_k(t)= b_k(0)e^{-i\Omega_k t}$ where $\Omega_k$ are the frequencies of the $\alpha-$FPUT chain, which are corrections to those of the harmonic chain. We get  corrections to the harmonic frequencies through the $\epsilon^2$ term in Eq.~\eqref{eq:evol2}, which is explained in more detail in Appendix.~\ref{app:freq}. We thus evolve $b_k$ and transform back to $a_k$, instead of evolving $a_k$ directly. We use the canonical transformation from $a_k$ to $b_k$ (Eq.~\eqref{eq:ct}), which we rewrite here after removing the $\delta$ terms:
\begin{equation}\label{eq:ct2}
	a_{k_1}=b_{k_1}+\epsilon\sum_{k_2}(A_{k_1,k_2,k_1-k_2}^{(1)}b_{k_2}b_{k_1-k_2}+A_{k_1,k_2,k_1+k_2}^{(2)}b_{k_2}^\star b_{k_1+k_2}+A_{k_1,k_2,-k_1-k_2}^{(3)}b_{k_2}^\star b_{-k_1-k_2}^\star)+O(\epsilon^2)~.
\end{equation}
Note that up to first order, we can invert the transformation Eq.~\eqref{eq:ct2} to:
\begin{equation}\label{eq:ct3}
	b_{k_1}=a_{k_1}-\epsilon\sum_{k_2}(A_{k_1,k_2,k_1-k_2}^{(1)}a_{k_2}a_{k_1-k_2}+A_{k_1,k_2,k_1+k_2}^{(2)}a_{k_2}^\star a_{k_1+k_2}+A_{k_1,k_2,-k_1-k_2}^{(3)}a_{k_2}^\star a_{-k_1-k_2}^\star)+O(\epsilon^2)~.
\end{equation}
 
We use the term perturbation theory to describe our method in this work because the construction of quasiperiodicity is being done by using the canonical transformation used to remove the non resonant interactions, which is a perturbation series. As mentioned in the introduction, we use the perturbation series up to first order in our computations (they remove only the three wave interactions), and discuss about higher order terms in Sec.~\ref{sec:Numerics}. We then compare the evolution of the normal mode energies $E_k = \omega_k\mid a_k\mid^2$ obtained by this method with that of molecular dynamics simulations obtained by numerically solving the Hamilton's equations. 

\section{Analytical solution of the quasiperiodicity}
\label{sec:Solution}
Let us construct an expression for quasiperiodicity in the $\alpha-$FPUT system for a generic system size $N$. We initially excite only the first and the last modes.
This is done in order to minimize the number of non zero terms in the sums of Eq.~\eqref{eq:ct2} and Eq.~\eqref{eq:ct3} (the authors of the original FPUT paper \cite{Fermi1955} excited only the first normal mode, and considered fixed boundary condition). So at $t=0$, $a_1(0)$ and $a_{-1}(0)$ are non zero (by $-k$, we mean the $(N-k)^{th}$ mode because of periodic boundary conditions) and the rest of them are zero. 
Let us now construct an expression for $a_k(t)$ up to first order. For that we have to first evaluate $b_k(0)$ using Eq.~\eqref{eq:ct3}. 

For $k_1=1$ one can notice that all the sums in Eq.~\eqref{eq:ct3} are zero since only $a_1$ and $a_{-1}$ are excited originally. For if $k_2 = 1$ then $k_1-k_2 =0$, which is the zero mode that is omitted from the dynamics. We also have $k_1+k_2 =2$ and $-k_1-k_2=-2$, both of which are not excited initially. 
Similarly for $k_2 = -1$ we have $k_1-k_2 =2$ (not excited initially), $k_1+k_2 =0$ and $-k_1-k_2=0$ (zero mode). We can do a similar analysis for $k_1=-1$ and infer that all the sums in Eq.~\eqref{eq:ct3} are zero. We thus have:
$b_1(0) = a_1(0)$, $b_{-1}(0)=a_{-1}(0)$. Similarly, one observes that $b_{k_1} = 0$ for all other values of $k_1$ except $k_1=2$ or $-2$. Note that this argument is valid only because we are computing $b_{k_1}$ up to first order. 

Now consider $k_1 = 2$. For $k_2 = 1$, it is possible to have $k_1-k_2 = 1$ but not $k_1 + k_2 =3$ and $-k_1-k_2 = -3$ (since they are not excited originally). For $k_2 = -1$ it is possible to have $k_1 + k_2 =1$ and $-k_1-k_2 = -1$ but not $k_1-k_2=3$. Thus, there will be three nonzero sums in Eq.~\eqref{eq:ct3} for $k_1=2$. In the same way, one can calculate that for $k_1=-2$, there are three nonzero sums in Eq.~\eqref{eq:ct3} $-$ for $k_2 = 1$, these are $-k_1-k_2=1$, $k_1+k_2=-1$ and for $k_2=-1$ it is $k_1-k_2=-1$. Thus, we get at $t=0$ the following expressions for $b_2$ and $b_{-2}$:
$$ b_2(0) \approx -\epsilon[A_{2,1,1}^{(1)}(a_1(0))^2+A_{2,-1,1}^{(2)}a_{-1}^\star(0) a_1(0)+A_{2,-1,-1}^{(3)}(a_{-1}^\star(0))^2]$$
$$ b_{-2}(0) \approx -\epsilon[A_{-2,-1,-1}^{(1)}(a_{-1}(0))^2+A_{-2,1,-1}^{(2)}a_{1}^\star(0) a_{-1}(0)+A_{-2,1,1}^{(3)}(a_{1}^\star(0))^2]$$
The purpose of computing $b_k(0)$ is two-fold. First, we can find the  frequencies of the $\alpha-$FPUT chain. The procedure to find corrections to the harmonic frequencies is explained in Appendix.~\ref{app:freq}. And second, we use $b_k(0)$ to find $a_k(t)$ through $b_k(t)$. Now we evolve $b_k$ in time as $b_k(t)=b_k(0)e^{-i\Omega_k t}$. Then using the canonical transformation Eq.~\eqref{eq:ct2} from $b_k$ to $a_k$, we get the evolution of $a_k$. Note that while transforming back to $a_k(t)$ we also have to consider the contributions of $b_2(t)$ and $b_{-2}(t)$ that are nonzero in addition to $b_1(t)$ and $b_{-1}(t)$. 
By doing a similar analysis that we have done in the previous paragraphs, we get the following expressions for the evolution of $a_k$ (accurate up to first order):
\begin{subequations}\label{eq:evolak}
\begin{equation}
\begin{split}
a_1(t) \approx a_1(0)e^{-i\Omega_1 t} + \epsilon [2A_{1,-1,2}^{(1)}a_{-1}(0) b_{2}(0) e^{-i(\Omega_1+\Omega_2)t}+2A_{1,1,-2}^{(3)}a_{1}^\star(0) b_{-2}^\star (0) e^{i(\Omega_1+\Omega_2)t}+\\A_{1,1,2}^{(2)}a_{1}^\star(0) b_{2}(0) e^{i(\Omega_1-\Omega_2)t}+A_{1,-2,-1}^{(2)}b_{-2}^\star(0) a_{-1}(0) e^{i(\Omega_2-\Omega_1)t}] 
\end{split}
\end{equation}
\begin{equation}
a_2(t) \approx b_2(0)e^{-i\Omega_2 t}+\epsilon[A_{2,1,1}^{(1)}a_{1}(0) a_{1}(0)e^{-2i\Omega_1t}+A_{2,-1,1}^{(2)}a_{-1}^\star(0) a_{1}(0)+A_{2,-1,-1}^{(3)}a_{-1}^\star(0) a_{-1}^\star(0)e^{2i\Omega_1t}]
\end{equation}
\begin{equation}
a_{-2}(t) \approx b_{-2}(0)e^{-i\Omega_2 t}+\epsilon[A_{-2,-1,-1}^{(1)}a_{-1}(0) a_{-1}(0)e^{-2i\Omega_1t}+A_{-2,1,-1}^{(2)}a_{1}^\star(0) a_{-1}(0)+A_{-2,1,1}^{(3)}a_{1}^\star(0) a_{1}^\star(0)e^{2i\Omega_1t}]
\end{equation}
\begin{equation}
\begin{split}
a_{-1}(t) \approx a_{-1}(0)e^{-i\Omega_1 t} + \epsilon [2A_{-1,1,-2}^{(1)}a_{1}(0) b_{-2}(0) e^{-i(\Omega_1+\Omega_2)t}+2A_{-1,-1,2}^{(3)}a_{-1}^\star(0) b_{2}^\star (0) e^{i(\Omega_1+\Omega_2)t}+\\A_{-1,-1,-2}^{(2)}a_{-1}^\star(0) b_{-2}(0) e^{i(\Omega_1-\Omega_2)t}+A_{-1,2,1}^{(2)}b_{2}^\star(0) a_{1}(0) e^{i(\Omega_2-\Omega_1)t}],~
\end{split} 
\end{equation}
\end{subequations}
where $\Omega_k$ are the frequencies of the $\alpha-$FPUT chain. 
These terms lead to the normal mode energies $E_k = \omega_k\mid a_k\mid^2$ with amplitude of order $\epsilon^2$ for $k=1,2,-1$ and $-2$. 
We don't mention other $a_k$ here because they are much smaller in magnitude. $E_3$ and $E_{-3}$ would have an amplitude of order $\epsilon^4$. Similarly $E_4$ and $E_{-4}$ would have an amplitude of order $\epsilon^6$, which are also not mentioned here. Note that we have to include terms up to order $\epsilon^2$ in Eqs.~\eqref{eq:ct2}, \eqref{eq:ct3} in order to get accurate expressions for $E_3$ and $E_{-3}$ (and also $E_4$ and $E_{-4}$), which lead to more complicated expressions for $a_k(t)$. 
Finally, note that the modes $k = 5,6,...$ and $k=-5,-6,...$ are not excited if we calculate Eqs.~\eqref{eq:ct2}, \eqref{eq:ct3} only up to first order. 

\section{Numerical results}
\label{sec:Numerics}
\begin{figure}[ht]
	\centering
	\hspace{-35mm}
	\includegraphics[width=0.5\textwidth]{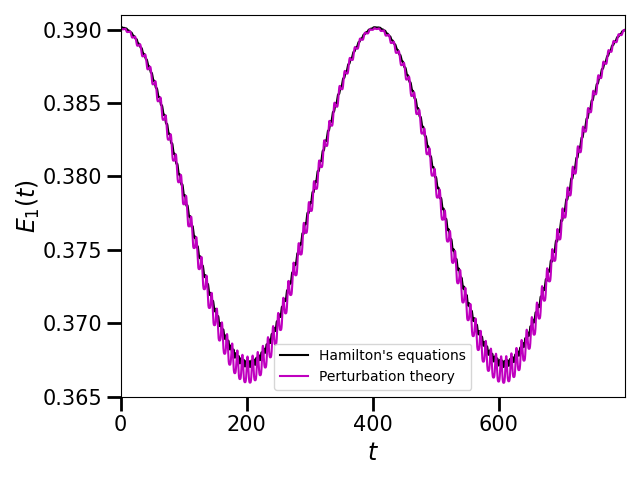}
	\put (-110,189) {$(a)$}
	\includegraphics[width=0.5\textwidth]{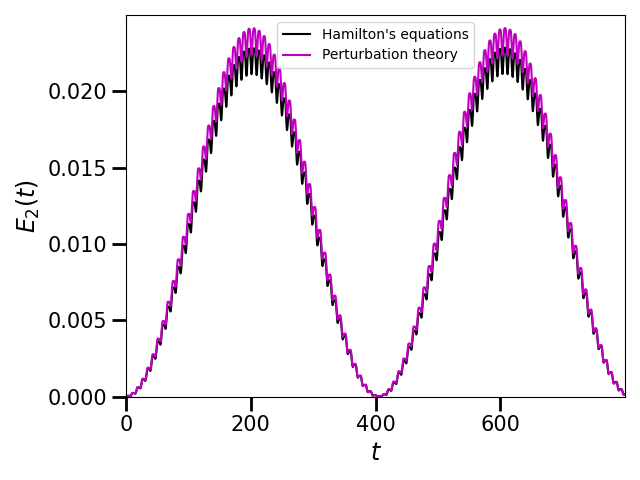}
	\put (-110,189) {$(b)$}
	\hspace{-40mm}
	\caption{Quasiperiodicity in the $\alpha-$FPUT chain: Plots show the time evolution of $E_1$ (left panel) and $E_2$ (right panel) starting from $a_1,a_{-1} = 1i$ and $a_k = 0$ for other values of $k$. Parameter values for this plot are $N=16$ and $\epsilon= 0.01$. The evolution of Hamilton's equations is plotted in black and the results of perturbation theory Eqs.~\eqref{eq:evolak} are shown in magenta. To a good approximation, $E_1$ and $E_2$ can be seen to oscillate with frequency $2\Omega_1-\Omega_2$.}
	\label{1601}
\end{figure}

We now discuss the numerical results. We initially excite the system to the first and the last normal modes of the harmonic chain (with amplitudes $a_1 = a_{-1} = 1i$ and $a_k = 0$ for other $k$) and then compare Eqs.~\eqref{eq:evolak} with the 
Hamilton's equations of motion of the $\alpha-$FPUT chain evolved by using a sixth order symplectic integrator \cite{Yoshida1990}. The time-step size is taken to be $0.01$. 
Fig.~\ref{1601} shows the evolution of the first two normal modes $E_1 = \omega_1\mid a_1\mid^2$ and $E_2= \omega_2\mid a_2\mid^2$ with system size $N=16$ and $\epsilon = 0.01$. The evolution of Hamilton's equations is plotted in black and the results of perturbation theory Eq.~\eqref{eq:evolak} are shown in magenta. We observe a good agreement between the two.

\subsection{Dependence on the frequency corrections}
\begin{figure}[ht]
	\centering
	\hspace{0mm}
	\includegraphics[width=0.5\textwidth]{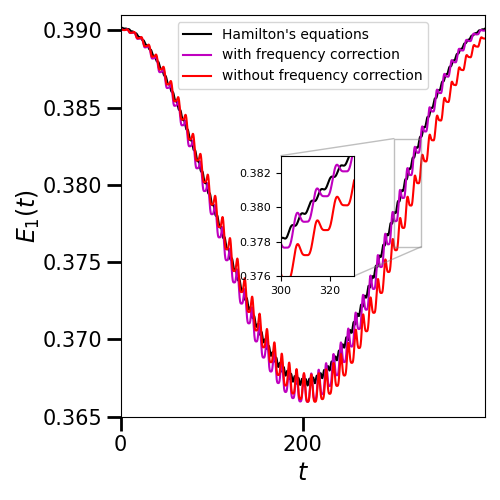}
	\put (-105,250) {$(a)$}
	\includegraphics[width=0.5\textwidth]{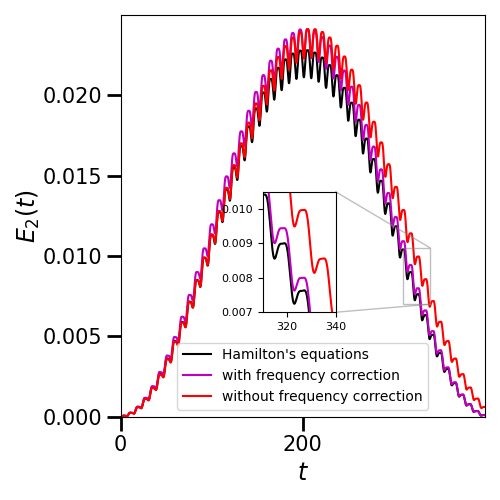}
	\put (-105,250) {$(b)$}
	\hspace{0mm}
	\caption{Dependence on the frequency correction: Same as Fig.~\ref{1601} but for shorter time. Here we illustrate the importance of the frequency corrections. The evolution of Hamilton's equations is plotted in black and the results of perturbation theory Eqs.~\eqref{eq:evolak} are shown in magenta. The result of perturbation theory Eqs.~\eqref{eq:evolak} when $\Omega_k$ is replaced by $\omega_k$ is plotted in red. The insets show the importance of computing the frequency corrections to the harmonic chain.}
	\label{frequencycorrection}
\end{figure}

We now demonstrate that it is indeed important to take the frequency corrections into account. We replot Fig.~\ref{1601} for shorter time in Fig.~\ref{frequencycorrection}. This time we also plot Eqs.~\eqref{eq:evolak} with $\Omega_k$ replaced by $\omega_k$, the harmonic frequencies. This is shown in red. The insets show that computing the frequency corrections lead to a noticeably better agreement between Hamiltonian evolution and perturbation theory. This is expected because if we don't take the frequency corrections into account then we get the same frequency of oscillations for all values of $\epsilon$ for which the wave turbulence formalism is applicable. 
 But we know that as $\epsilon$ is decreased the quasiperiodic frequency should reduce, and so the time period should increase. This is captured by the frequency corrections and so we replace $\omega_k$ with $\Omega_k$. The procedure to calculate the frequency corrections is described in  Appendix.~\ref{app:freq}. 
 A more serious problem that would arise by not including the frequency corrections is that these terms that are not included in the frequency correction would then contribute to the dynamics, and these lead to a secular growth as described in \cite{Zakharov1999,Nazarenko2011,Onorato2020}. In the standard perturbation theory such a technique is known as the Poincare-Lindstedt method. 
 
\subsection{Dependence on the nonlinearity  parameter}
\begin{figure}[ht]
	\centering
	\hspace{-35mm}
	\includegraphics[width=0.5\textwidth]{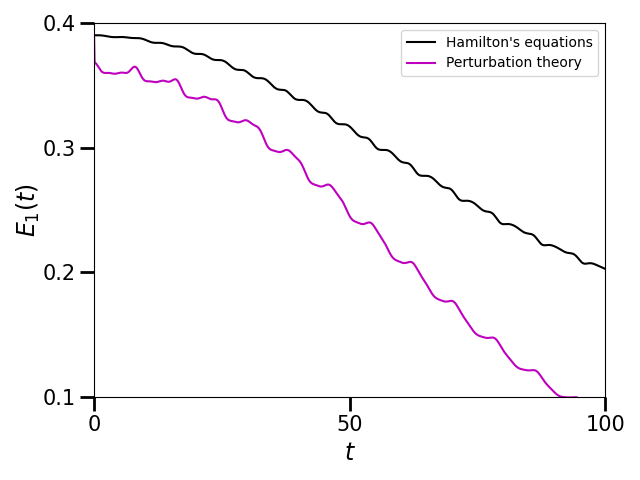}
	\put (-110,189) {$(a)$}
	\includegraphics[width=0.5\textwidth]{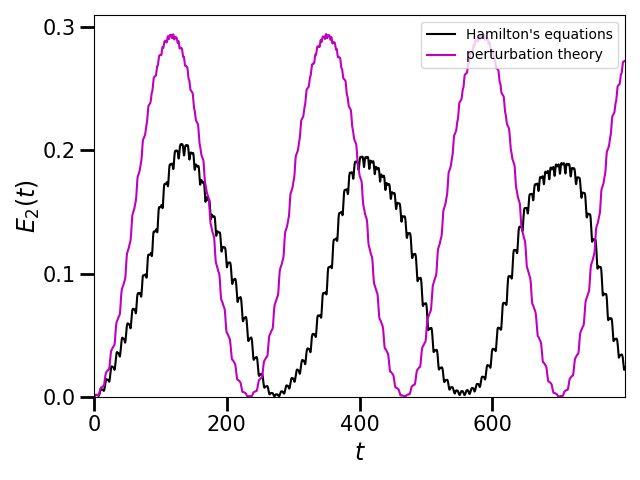}
	\put (-110,189) {$(b)$}
	\hspace{-40mm}
	\caption{Dependence on the nonlinearity parameter $\epsilon$: 
	Plots show the time evolution of $E_1$ (left panel) and $E_2$ (right panel) for $\epsilon = 0.05$ starting from the same initial condition as Fig.~\ref{1601} with $N = 16$. We now observe a breakdown of agreement between Hamiltonian dynamics and perturbation theory for $E_1$ from early times, and also a poorer agreement in the evolution of $E_2$.}
	\label{16particles}
\end{figure}

We now move on to study how our method captures the quasiperiodic behaviour for larger values of $\epsilon$. This is shown in Fig.~\ref{16particles} where we plot the evolution of $E_1$ on the left panel and $E_2$ on the right for $N=16$ and $\epsilon = 0.05$. We see that the agreement between Hamiltonian dynamics and perturbation theory is poor now. In fact, there is a breakdown of agreement for $E_1$ even at very small times. This is not expected to happen for $\epsilon = 0.05$ as $0.05$ is still a small perturbation. 
We show in Sec.~\ref{ssec:Math} that the presence of small denominators in Eqs.~\eqref{eq:ct2},~\eqref{eq:ct3} would lead to further restrictions on $\epsilon$.
\subsection{Dependence on the system size}
\begin{figure}[ht]
	\centering
	\hspace{-35mm}
	\includegraphics[width=0.5\textwidth]{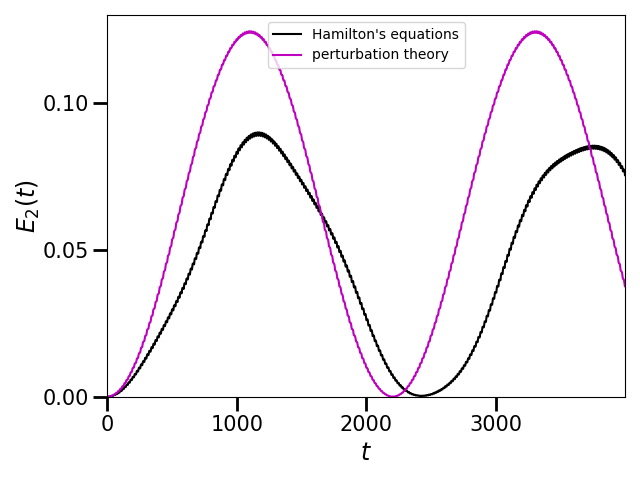}
	\put (-110,189) {$(a)$}
	\includegraphics[width=0.5\textwidth]{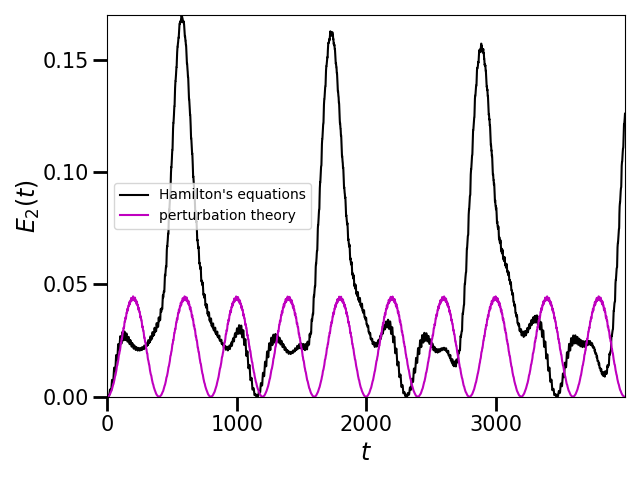}
	\put (-110,189) {$(b)$}
	\hspace{-40mm}
	\caption{Dependence on the system size $N$: Plots show the time evolution of $E_2$ for $\epsilon = 0.01$ (left panel) and $\epsilon = 0.05$ (right panel) starting from the same initial condition as Fig.~\ref{1601} for $N=32$. We now observe a poorer agreement  even for $\epsilon = 0.01$. From the right panel we see that the quasiperiodic behaviour in the $\alpha-$FPUT chain that is observed in Fig.~\ref{FPU} is not at all captured by perturbation theory.}
	\label{32particles}
\end{figure}

We now study how our method captures the quasiperiodic behaviour for larger system size. Fig.~\ref{32particles} shows the time evolution of $E_2$ for $\epsilon = 0.01$ (left panel) and $\epsilon = 0.05$ (right panel) for $N=32$ starting from the same initial condition as Fig.~\ref{1601}. We see that $\epsilon = 0.01$ is also not small enough to get a good agreement for $N=32$. From the right panel we see that the quasiperiodic behaviour (Fig.~\ref{FPU}) in the $\alpha-$FPUT chain that resembles the plot in the original paper \cite{Fermi1955} is not captured by perturbation theory. We now explain this lack of agreement by pointing out that the small denominators in Eqs.~\eqref{eq:ct2},~\eqref{eq:ct3} will become more pronounced as the system size is increased.

\subsection{Evidence of divergence in the canonical transformation}
\label{ssec:Math}
We now relate the observed discrepancies in the agreement between perturbation theory and Hamilton's equations to the presence of small denominators in the canonical transformations Eqs.~\eqref{eq:ct2},~\eqref{eq:ct3}.
For the initial conditions that we have considered, one can observe that as the system size is increased, terms like $\epsilon A_{1,1,2}^{(2)}$ in Eqs.~\eqref{eq:ct2},~\eqref{eq:ct3} and the solution Eqs.~\eqref{eq:evolak} become large even for smaller $\epsilon$ because of the small denominator in $A_{1,1,2}^{(2)}$, which has the form $2sin(\pi/N)-sin(2\pi/N)$. This implies that the higher order terms in Eqs.~\eqref{eq:ct2},~\eqref{eq:ct3}, which we have neglected while deriving Eqs.~\eqref{eq:evolak}, become non-negligible when compared to (sometimes even larger than) the lower order terms. This can clearly be noticed by considering higher order terms in Eqs.~\eqref{eq:ct2}. These terms are of the form $\epsilon^2 \sum B_{k_1,k_2,k_3,k_4}b_{k_2}^\star b_{k_3}b_{k_4}\delta_{k_1+k_2,k_3+k_4} $, where $ B_{k_1,k_2,k_3,k_4}$ is given by \cite{Dyachenko1994}:
\begin{equation}\label{eq:B}
\begin{split}
B_{k_1,k_2,k_3,k_4} = A_{k_2,k_3,k_2-k_3}^{(1)}A_{k_4,k_1,k_4-k_1}^{(1)}+ A_{k_2,k_4,k_2-k_4}^{(1)}A_{k_3,k_1,k_3-k_1}^{(1)}- A_{k_1,k_3,k_1-k_3}^{(1)}A_{k_4,k_2,k_4-k_2}^{(1)}- A_{k_2,k_4,k_2-k_4}^{(1)}A_{k_3,k_2,k_3-k_2}^{(1)}\\- A_{k_1+k_2,k_1,k_2}^{(1)}A_{k_3+k_4,k_3,k_4}^{(1)}+ A_{-k_1-k_2,k_1,k_2}^{(3)}A_{-k_3-k_4,k_3,k_4}^{(3)}
\end{split}	
\end{equation}
For $(k_1,k_2,k_3,k_4)=(1,2,1,2)$ we can see that the first term in Eq.~\eqref{eq:B} leads to a term $\sim(\epsilon A_{2,1,1}^{(1)})^2$, which is the square of a small denominator. This indicates that there could be divergences in the canonical transformation, and a possible breakdown of perturbation theory.  
Thus, we argue that the divergences in the canonical transformation can be avoided only if Eqs.~\eqref{eq:ct2},~\eqref{eq:ct3} remain bounded. This is possible if $\epsilon A_{1,1,2}^{(2)}\ll1$. Hence the perturbation series Eqs.~\eqref{eq:ct2},~\eqref{eq:ct3} are expected to converge only if $\epsilon A_{1,1,2}^{(2)}\ll1$. We point out that there will be a good agreement between the evolution of the $\alpha-$FPUT chain obtained by perturbation theory and by evolving Hamilton's equations (such as in Fig.~\ref{1601}) when $\epsilon$ is small enough so that $\epsilon A_{1,1,2}^{(2)}\ll 1$. 
Thus for system size $N$ we expect Eqs.~\eqref{eq:ct2},~\eqref{eq:ct3} to remain valid only if:
\begin{equation}\label{eq:breakdown}
	\epsilon\frac{\sqrt{sin^2(\pi/N)sin(2\pi/N)}}{2sin(\pi/N)-sin(2\pi/N)}\ll1~.
\end{equation}
For $N=8$ the limitation becomes $\epsilon\ll0.18$, while for $N=16$ this becomes $\epsilon\ll0.06$ and for $N=32$ it becomes $\epsilon\ll0.02$. As the system size $N$ increases Eqs.~\eqref{eq:ct2},~\eqref{eq:ct3} are expected to converge only if $\epsilon \ll O(N^{-1.5})$. Hence we have to move on to smaller and smaller $\epsilon$ in the thermodynamic limit in order for Eqs.~\eqref{eq:ct2},~\eqref{eq:ct3} to be convergent. Otherwise the higher order terms would be of the order of (or even larger than) the lower order terms, and perturbation theory is expected to break down. 
Note that terms such as $A_{1,-1,2}^{(1)}$ and $A_{1,1,-2}^{(3)}$ do not pose any further restrictions on the nonlinearity parameter because the denominators are not small enough when compared to that of $A_{1,1,2}^{(2)}$. 

\begin{figure}[ht]
	\centering
	\hspace{-35mm}
	\includegraphics[width=0.5\textwidth]{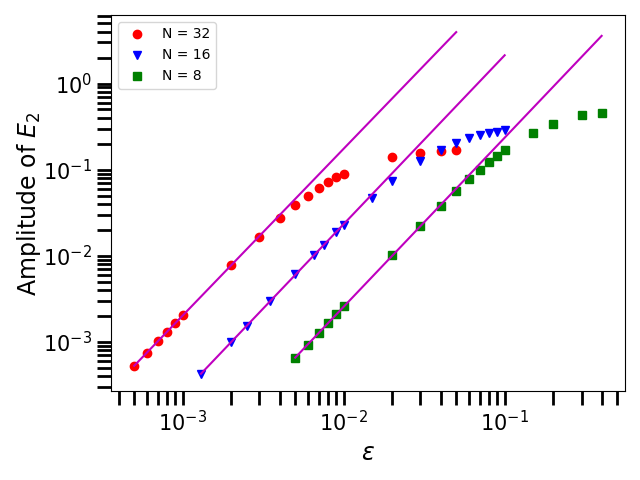}
	\put (-110,189) {$(a)$}
	\includegraphics[width=0.5\textwidth]{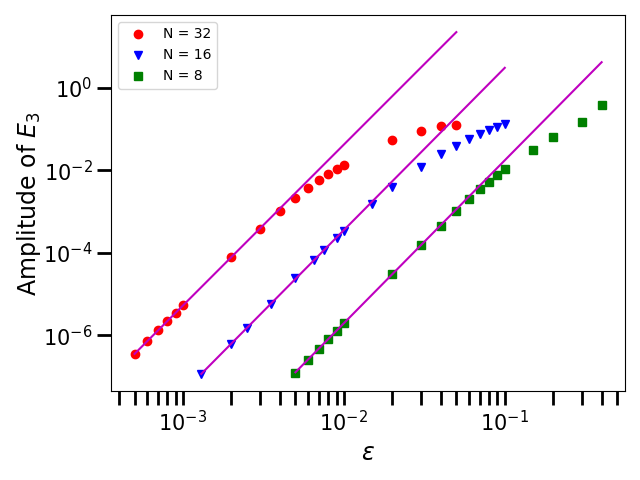}
	\put (-110,189) {$(b)$}
	\hspace{-40mm}
	\caption{Breakdown of perturbation theory for different $N$: Here we plot the amplitude of $E_2$ (left panel) and $E_3$ (right panel) at very short times for different $\epsilon$ after numerically solving the Hamilton's equations of motion starting from the same initial condition as Fig.~\ref{1601}. The slope of the lines in the left panel is $2$, while it is $4$ in the right panel. 
	We can clearly see that the parameters $N=32$ and $\epsilon=0.05$ in Fig.~\ref{FPU} correspond to the strong nonlinearity regime.}
	\label{Breakdown}
\end{figure}

Even though we have given indications that the perturbation series Eqs.~\eqref{eq:ct2},~\eqref{eq:ct3} diverge if $\epsilon A_{1,1,2}^{(2)}$ is large enough, we have not proved it mathematically. This is beyond the scope of this study. It may as well be argued that all the small denominators cancel in all the higher order terms, and we end up with a convergent series whenever $\epsilon \ll 1$. We now show numerical evidence that this is not the case, and that the small denominators in the higher order do not cancel. We plot in Fig.~\ref{Breakdown} the amplitude of oscillations of $E_2$ and $E_3$ at very short times as a function of $\epsilon$ after evolving the $\alpha-$FPUT system by using a symplectic integrator. We have considered the same initial condition as our other plots. 
Here are some of the interesting observations:
\begin{itemize}
	\item [--] From the left (right) panel we observe that for small values of $\epsilon$ the amplitudes of $E_2$ ($E_3$) fit to a straight line of slope $2$ ($4$) on a log-log plot, confirming our calculations (predictions) of the $\epsilon^2$ ($\epsilon^4$) dependence of the amplitude of $E_2$ ($E_3$ respectively). 
	\item[--] We also find from both the panels that for larger $\epsilon$ the amplitudes deviate from the straight line showing clearly where the higher order terms are not negligible. Deviation from straight lines start to become noticeable when $\epsilon A_{1,1,2}^{(2)}$ is comparable to $1$, providing evidence that small denominators in Eqs.~\eqref{eq:ct2},~\eqref{eq:ct3} do not cancel in higher order terms. Had the terms with small denominators been cancelled, the amplitudes would have scaled as $\epsilon^{2}$ for $E_2$ and $\epsilon^{4}$ for $E_3$ even for those $\epsilon\ll1$ for which $\epsilon A_{1,1,2}^{(2)}$ is large enough. 
	\item[--] And finally note that as the system size is increased the breakdown in agreement is observed at a lower $\epsilon$, consistent with our estimates. 
	
\end{itemize}
This deviation from the straight line shows evidence that for $\epsilon\ll1$ such that $\epsilon A_{1,1,2}^{(2)}$ is large enough, the contribution of the higher order terms in the canonical transformations Eqs.~\eqref{eq:ct2},~\eqref{eq:ct3} is as large as (or even larger than) the lower order terms. This leads to the divergence of the canonical transformations. Thus, we provide evidence of a strong nonlinearity regime where perturbation theory cannot be applied to remove the three wave interactions, and we cannot arrive at Eq.~\eqref{eq:evol2}, which is the starting point of wave turbulence.  We see that we have not escaped the problem of small divisors completely yet. Interestingly, some comments on the mathematical challenges of wave turbulence when applied to the $\alpha-$FPUT system have been made in some of the earlier works done on the subject (for instance \cite{Kramer2002,Biello2002}).


Finally, note from Fig.~\ref{Breakdown} that $\epsilon = 0.05$ is in the strong nonlinearity regime for $N=32$ (shown in Fig.~\ref{FPU}, which was observed in the original paper \cite{Fermi1955}). In the weak nonlinearity regime where perturbation theory is applicable, the modes $k=1$ and $-1$ excited initially can only transfer energy up to four modes, and the fifth mode cannot get excited (as we found in Sec.~\ref{sec:Solution}). However, the fifth mode is found to have energy (as shown in Fig.~\ref{FPU}) and this gives us the initial clue that we may be in the strong nonlinearity regime for $N=32$ and $\epsilon = 0.05$.



\section{Conclusion}
\label{sec:Conclusion}
In this work we tried to reconstruct the quasiperiodic behaviour observed in the weakly nonlinear $\alpha-$FPUT system (such as the one in Fig.~\ref{FPU}, which resembles the plot observed in the original paper \cite{Fermi1955}) using the canonical transformation Eq.~\eqref{eq:ct2} used to remove the three wave interactions. We obtained analytical expressions Eqs.~\eqref{eq:evolak} of the evolution of $a_k$ (given by $a_k = \frac{1}{\sqrt{2\omega_k}}(P_k-i\omega_kQ_k)$) when only $a_1$ and $a_{-1}$ are excited initially. We also used the wave turbulence formalism to compute the frequencies of the $\alpha-$FPUT chain, which are  perturbative corrections to that of the harmonic chain, in order to construct the evolution even more accurately. We then compared our results with that of the molecular dynamics simulations obtained by numerically solving the Hamilton's equations of motion.
Our work can be summarized by linking the following observations:
\begin{enumerate}
	\item For the initial conditions that we have considered, we find that the agreement between Eqs.~\eqref{eq:evolak} and Hamiltonian dynamics is good only for $\epsilon A_{1,1,2}^{(2)}\ll1$. This leads to $\epsilon \ll O(N^{-1.5})$, where $N$ is the system size.
	\item There is a small denominator in $A_{1,1,2}^{(2)}$, that is more pronounced as the system size increases. We show that small denominators are also present in higher order terms in the canonical transformation Eq.~\eqref{eq:ct2}, and perturbation theory is expected to be valid only when  $\epsilon \ll O(N^{-1.5})$. 
	\item Using the molecular dynamics simulations we plotted the amplitude of $E_k = \omega_k\mid a_k\mid^2$ 
	as a function of $\epsilon$ for $k=2,3$ at very short times (Fig.~\ref{Breakdown}). We find that there is a deviation from the straight line that is noticeable when $\epsilon A_{1,1,2}^{(2)}$ is comparable to $1$, which demonstrates that the small denominators do not cancel each other in higher orders, which makes a strong point for our claim that the canonical transformation Eq.~\eqref{eq:ct2} diverges, and we are in the strong nonlinearity regime.
	\item The quasiperiodic behaviour in Fig.~\ref{FPU}, which resembles the plot observed in the original paper \cite{Fermi1955} corresponds to the strong nonlinearity regime where the small denominators do not cancel each other in higher orders. 

\end{enumerate}


An interesting extension of this work is to understand why the strong nonlinearity is affecting the quasiperiodic behaviour but not speeding up the rate of thermalization (we should expect faster than the wave turbulence prediction of $1/\epsilon^8$ dependence of the equilibration time). This may be due to finite size effects and has to be studied further. One way to explain the disagreement in the observed quasiperiodicity is to capture the reversible dynamics by using the ideas of \cite{Guasoni2017}, which uses a nonequilibrium spatiotemporal kinetic formulation that accounts for the existence of phase correlations among incoherent waves. This has to be investigated further, and is beyond the scope of this study. There have been some studies done recently that describe the FPUT system as a perturbation of the Toda system (exponential interaction between the nearest neighbours) \cite{Toda1967,Toda1975,Goldfriend2019,Fu2019}. An interesting question would be to link this description with the wave turbulence formalism.
 
Earlier studies \cite{Onorato2015} have predicted that the equilibration time $\tau$ scales as $1/\epsilon^8$ (for finite system sizes), which was found to happen (numerically) when $\epsilon A_{1,1,2}^{(2)}$ is comparable to (sometimes larger than) $1$. But we have pointed out in our study that there is a divergence in the canonical transformation used to remove the three wave interactions when $\epsilon A_{1,1,2}^{(2)}$ is comparable to $1$ even for small $\epsilon$. 
We thus point out that there are mathematical challenges when the wave turbulence formalism is applied to the  $\alpha-$FPUT system. It is possible that if more number of modes are excited initially then these small denominators would cancel each other in all the higher orders, and we end up with a convergent series whenever $\epsilon\ll 1$. However, that brings us back to the question of whether there is a stochasticity threshold in the system \cite{Chirikov1960, Izrailev1966,Livi1985,Deluca1995,Casetti1996}. It may as well be argued that if only the first and the last modes are excited initially then the energy may not reach the higher energy modes (as we found in Sec.~\ref{sec:Solution}) and eventually we may not observe equipartition of energy, unless there are small denominators. This leads to an apparent conclusion that there exists a stochasticity threshold in the system. This is an interesting possibility which has to be investigated using further mathematical and numerical studies, and is beyond the scope of this study. Nevertheless, our work shows that the problem of thermalization in the $\alpha-$FPUT problem still retains an element of ``surprise''. 
\section*{Acknowledgments} I acknowledge Abhishek Dhar, Amirali Hannani, Amit Apte, Kabir Ramola, Miguel Onorato, Varun Dubey and Wojciech De Roeck for helpful discussions at various stages of this work. I also thank Miguel Onorato for the hospitality at the Summer school on Wave turbulence and beyond in Turin, Italy. I acknowledge the grant $G098919N$ from the Research Foundation – Flanders (FWO), Belgium and also the Department of Atomic Energy, Government of India for funding my positions during which this work has been done.
\begin{appendix}
\section{Frequency corrections}
\label{app:freq}
Let us consider the evolution equation Eq.~\eqref{eq:evol2}:
$$i\frac{\partial b_{k_1}}{\partial t} = \omega_{k_1} b_{k_1} + \epsilon^2 \sum_{k_2,k_3,k_4} T_{k_1,k_2,k_3,k_4} b_{k_2}^\star b_{k_3}b_{k_4}\delta_{k_1+k_2,k_3+k_4} +O(\epsilon^3)~.$$
The sum in the above equation can be broken into $2$ parts. The first sum corresponds to $k_1 = k_3$ and $k_2 = k_4$. The second sum corresponds to the remaining terms. The equilibration problem in the $\alpha-$FPUT chain has been discussed in terms of the second sum, which includes the four wave resonances. For the purpose of computing the frequency correction however, we just need the first part. We thus write Eq.~\eqref{eq:evol2} as:
\begin{equation}\label{eq:evol3}
i\frac{\partial b_{k_1}}{\partial t} = \omega_{k_1}b_{k_1}+ 2\epsilon^2 \sum_{k_2} T_{k_1,k_2,k_1,k_2} b_{k_2}^\star b_{k_2}b_{k_1}-\epsilon^2 T_{k_1,k_1,k_1,k_1} b_{k_1}^\star b_{k_1}b_{k_1} + \text{second sum} +O(\epsilon^3)~.
\end{equation}
The factor $2$ accounts for the permutation $k_1 = k_4$ and $k_2 = k_3$. Since $k_1 = k_2 = k_3 = k_4$ has no permutation, the factor of $2$ should not be included for this term and hence we have the subtracted term. Thus, we get the second order correction to the frequencies by writing Eq.~\eqref{eq:evol3} as:
\begin{equation}\label{eq:evol4}
i\frac{\partial b_{k_1}}{\partial t} \approx (\omega_{k_1}+ 2\epsilon^2 \sum_{k_2} T_{k_1,k_2,k_1,k_2} b_{k_2}^\star b_{k_2}-\epsilon^2 T_{k_1,k_1,k_1,k_1} b_{k_1}^\star b_{k_1})b_{k_1}~.
\end{equation}
The corrected frequencies $\Omega$ are given by:
\begin{equation}\label{eq:omegapert}
	\Omega_{k_1} = \omega_{k_1}+\epsilon^2 (2\sum_{k_2} T_{k_1,k_2,k_1,k_2} b_{k_2}^\star b_{k_2}- T_{k_1,k_1,k_1,k_1} b_{k_1}^\star b_{k_1})+O(\epsilon^3)~,
\end{equation}
where the $b_k$ are computed at $t=0$. In this way, $i\frac{\partial b_{k_1}}{\partial t} \approx \Omega_{k_1} b_{k_1}$ is a better approximation to the evolution equation when compared to $i\frac{\partial b_{k_1}}{\partial t} \approx \omega_{k_1} b_{k_1}-$  the former takes into account the frequency corrections to the $\alpha-$FPUT chain due to its nonlinearity. Note that for the $\alpha-FPUT$ chain the first order correction to the harmonic frequencies is zero. In order to compute higher order corrections to the frequencies (the next term turns out to be of order $\epsilon^4$ rather than $\epsilon^3$), we have to similarly split the higher order sums into two. 
Note that the terms responsible for the frequency correction are not included in the subsequent dynamics. If however, we don't normalize the frequencies and instead, use these terms while computing the evolution, then there will be secular terms in equation Eq.~\eqref{eq:evol3} and the formalism of wave turbulence will not be self-consistent \cite{Zakharov1999, Nazarenko2011}. 

The general expression for $T_{k_1,k_2,k_3,k_4}$ depends on the transfer matrix $V_{k_1,k_2,k_3}$ (for the $\alpha-$FPUT system $V_{k_1,k_2,k_3}$ is defined in Eq.~\eqref{eq:transfer}) and is given in \cite{Dyachenko1994}. We reproduce the expression here:
\begin{equation}\label{eq:T1234}
\begin{split}
T_{k_1,k_2,k_3,k_4} = -V_{k_1,k_3,k_1-k_3}V_{k_4,k_2,k_4-k_2}[\frac{1}{\omega_{k_3}+\omega_{k_1-k_3}-\omega_{k_1}}+\frac{1}{\omega_{k_2}+\omega_{k_4-k_2}-\omega_{k_4}}]\\-V_{k_2,k_3,k_2-k_3}V_{k_4,k_1,k_4-k_1}[\frac{1}{\omega_{k_3}+\omega_{k_2-k_3}-\omega_{k_2}}+\frac{1}{\omega_{k_1}+\omega_{k_4-k_1}-\omega_{k_4}}]\\-V_{k_1,k_4,k_1-k_4}V_{k_3,k_2,k_3-k_2}[\frac{1}{\omega_{k_4}+\omega_{k_1-k_4}-\omega_{k_1}}+\frac{1}{\omega_{k_2}+\omega_{k_3-k_2}-\omega_{k_3}}]\\-V_{k_2,k_4,k_2-k_4}V_{k_3,k_1,k_3-k_1}[\frac{1}{\omega_{k_4}+\omega_{k_2-k_4}-\omega_{k_2}}+\frac{1}{\omega_{k_1}+\omega_{k_3-k_1}-\omega_{k_3}}]\\-V_{k_1+k_2,k_1,k_2}V_{k_3+k_4,k_3,k_4}[\frac{1}{\omega_{k_1+k_2}-\omega_{k_1}-\omega_{k_2}}+\frac{1}{\omega_{k_3+k_4}-\omega_{k_3}-\omega_{k_4}}]\\-V_{-k_1-k_2,k_1,k_2}V_{-k_3-k_4,k_3,k_4}[\frac{1}{\omega_{k_1+k_2}+\omega_{k_1}+\omega_{k_2}}+\frac{1}{\omega_{k_3+k_4}+\omega_{k_3}+\omega_{k_4}}] ~.
\end{split}
\end{equation}
From this we get the following expression for $T_{k_1,k_2,k_1,k_2}$:
\begin{equation}\label{eq:T1212}
\begin{split}
T_{k_1,k_2,k_1,k_2} = -\frac{2(V_{k_2,k_1,k_2-k_1})^2}{\omega_{k_1}+\omega_{k_2-k_1}-\omega_{k_2}} -\frac{2(V_{k_1,k_2,k_1-k_2})^2}{\omega_{k_2}+\omega_{k_1-k_2}-\omega_{k_1}} -\frac{2(V_{k_1+k_2,k_1,k_2})^2}{\omega_{k_1+k_2}-\omega_{k_1}-\omega_{k_2}}
-\frac{2(V_{-k_1-k_2,k_1,k_2})^2}{\omega_{k_1+k_2}+\omega_{k_1}+\omega_{k_2}} ~.
\end{split}
\end{equation}
Note that the zero mode does not participate in the dynamics and hence, there are no zero denominators in the expression for $T_{k_1,k_1,k_1,k_1}$. Using the expression for $T_{k_1,k_2,k_1,k_2}$ in Eq.~\eqref{eq:omegapert} we thus compute the frequencies of the $\alpha-$FPUT system.
\end{appendix}

\bibliographystyle{unsrt}
\bibliography{references}

\begin{thebibliography}{10}

\bibitem{Fermi1955}
E.~Fermi, J.~Pasta, and S.~Ulam.
\newblock Studies of the nonlinear problems, (1955).
\newblock Los Alamos National Laboratory Report (LA1940), also in Collected
  Papers of Enrico Fermi 2.

\bibitem{dauxois2008fermi}
T.~Dauxois.
\newblock Fermi, {P}asta, {U}lam, and a mysterious lady.
\newblock {\em Physics today}, 61(1):55, (2008).

\bibitem{Zabusky1965}
N.~J. {Zabusky} and M.~D. {Kruskal}.
\newblock {Interaction of ``Solitons'' in a Collisionless Plasma and the
  Recurrence of Initial States}.
\newblock {\em Phys. Rev. Lett.}, 15:240--243, (1965).

\bibitem{Chirikov1960}
B.~V. Chirikov.
\newblock Resonance processes in magnetic traps.
\newblock {\em The Soviet Journal of Atomic Energy}, 6(6):464--470, (1960).

\bibitem{Izrailev1966}
F.~M. Izrailev and B.~V. Chirikov.
\newblock {Statistical properties of a nonlinear string}.
\newblock {\em Dokl. Akad. Nauk SSSR}, 166(1):57--59, (1966).

\bibitem{Casetti1996}
L.~Casetti, M.~Cerruti-Sola, M.~Pettini, and E.~Cohen.
\newblock The {F}ermi-{P}asta-{U}lam problem revisited: Stochasticity
  thresholds in nonlinear {H}amiltonian systems.
\newblock {\em Phys. Rev. E}, 55(6), (1996).

\bibitem{Benettin2018}
G.~Benettin, S.~Pasquali, and A.~Ponno.
\newblock The {F}ermi–{P}asta–{U}lam problem and its underlying integrable
  dynamics: An approach through {L}yapunov exponents.
\newblock {\em Journal of Statistical Physics}, 171(4):521–542, Mar (2018).

\bibitem{Liu2021}
Y.~Liu and D.~He.
\newblock Analytical approach to lyapunov time: Universal scaling and
  thermalization.
\newblock {\em Physical Review E}, 103(4):L040203, (2021).

\bibitem{Marin1996}
J.~Marin and S.~Aubry.
\newblock Breathers in nonlinear lattices: numerical calculation from the
  anticontinuous limit.
\newblock {\em Nonlinearity}, 9:1501--1528, (1996).

\bibitem{Flach2005}
S.~Flach, M.~V. Ivanchenko, and O.~I. Kanakov.
\newblock $q$-{B}reathers and the {F}ermi-{P}asta-{U}lam problem.
\newblock {\em Phys. Rev. Lett.}, 95(6):064102, (2005).

\bibitem{Flach2006}
S.~Flach, M.~V. Ivanchenko, and O.~I. Kanakov.
\newblock $q$-breathers in {F}ermi-{P}asta-{U}lam chains: Existence,
  localization, and stability.
\newblock {\em Phys. Rev. E}, 73(3):036618, (2006).

\bibitem{Danieli2017}
C.~Danieli, D.~K. Campbell, and S.~Flach.
\newblock Intermittent many-body dynamics at equilibrium.
\newblock {\em Phys. Rev. E}, 95(6):060202, (2017).

\bibitem{Ganapa2020}
S.~Ganapa, A.~Apte, and A.~Dhar.
\newblock Thermalization of local observables in the $\alpha$-{FPUT} chain.
\newblock {\em Journal of Statistical Physics}, 180(1):1010--1030, (2020).

\bibitem{Zakharov1992}
V.~E. Zakharov V.~S. L{\textquotesingle}vov and G.~Falkovich.
\newblock Statistical description of weak wave turbulence.
\newblock In {\em Kolmogorov Spectra of Turbulence I: Wave Turbulence}, pages
  63--82. Springer Berlin Heidelberg, (1992).

\bibitem{Nazarenko2011}
S.~Nazarenko.
\newblock {\em Wave Turbulence}.
\newblock Springer Berlin Heidelberg, (2011).

\bibitem{Onorato2015}
M.~Onorato, L.~Vozella, D.~Proment, and Y.V. Lvov.
\newblock Route to thermalization in the
  $\alpha$-{F}ermi{\textendash}{P}asta{\textendash}{U}lam system.
\newblock {\em PNAS}, 112(14):4208--4213, (2015).

\bibitem{Lvov2018}
Y.~V. Lvov and M.~Onorato.
\newblock Double scaling in the relaxation time in the
  $\beta$-{F}ermi-{P}asta-{U}lam-{T}singou model.
\newblock {\em Phys. Rev. Lett.}, 120:144301, Apr (2018).

\bibitem{Pistone2019}
L.~Pistone, S.~Chibbaro, M.~Bustamante, Y.~L'vov, and M.~Onorato.
\newblock Universal route to thermalization in weakly-nonlinear one-dimensional
  chains.
\newblock {\em Mathematics in Engineering}, 1(4):672, (2019).

\bibitem{Bustamante2019}
M.D. Bustamante, K.~Hutchinson, Y.V. Lvov, and M.~Onorato.
\newblock Exact discrete resonances in the {F}ermi-{P}asta-{U}lam–{T}singou
  system.
\newblock {\em Communications in Nonlinear Science and Numerical Simulation},
  73:437--471, (2019).

\bibitem{Ford1992}
J.~Ford.
\newblock The {F}ermi-{P}asta-{U}lam problem: Paradox turns discovery.
\newblock {\em Physics Reports}, 213(5):271 -- 310, (1992).

\bibitem{Weissert1997}
T.~P. Weissert.
\newblock {\em The genesis of simulation in dynamics: pursuing the
  Fermi-Pasta-Ulam problem}.
\newblock Springer, (2012).

\bibitem{Berman2005}
G.~P. Berman and F.~M. Izrailev.
\newblock The {F}ermi-{P}asta-{U}lam problem: Fifty years of progress.
\newblock {\em Chaos: An Interdisciplinary Journal of Nonlinear Science},
  15(1):15104, (2005).

\bibitem{Gallavotti2008}
G.~Gallavotti.
\newblock Introduction to {F}{P}{U}.
\newblock In Giovanni Gallavotti, editor, {\em The Fermi-Pasta-Ulam Problem: A
  Status Report}. Springer Berlin Heidelberg, (2008).

\bibitem{Benettin2013}
G.~{Benettin}, H.~{Christodoulidi}, and A.~{Ponno}.
\newblock {The Fermi-Pasta-Ulam Problem and Its Underlying Integrable
  Dynamics}.
\newblock {\em Journal of Statistical Physics}, 152:195--212, Jul (2013).

\bibitem{Christodoulidi2010}
H.~Christodoulidi, C.~Efthymiopoulos, and T.~Bountis.
\newblock Energy localization on $q$-tori, long-term stability, and the
  interpretation of {F}ermi-{P}asta-{U}lam recurrences.
\newblock {\em Phys. Rev. E}, 81(1):016210, (2010).

\bibitem{Livi1985}
R.~Livi, M.~Pettini, S.~Ruffo, M.~Sparpaglione, and A.~Vulpiani.
\newblock Equipartition threshold in nonlinear large {H}amiltonian systems: The
  {F}ermi-{P}asta-{U}lam model.
\newblock {\em Phys. Rev. A}, 31(2):1039--1045, (1985).

\bibitem{Deluca1995}
J.~{Deluca}, A.~J. {Lichtenberg}, and S.~{Ruffo}.
\newblock {Energy transitions and time scales to equipartition in the
  Fermi-Pasta-Ulam oscillator chain}.
\newblock {\em Phys. Rev. E}, 51(4):2877--2885, (1995).

\bibitem{Dyachenko1994}
A.~I. Dyachenko and V.~E. Zakharov.
\newblock Is free-surface hydrodynamics an integrable system?
\newblock {\em Physics Letters A}, 190(2):144--148, (1994).

\bibitem{Dyachenko1995}
A.I. Dyachenko, Y.V. Lvov, and V.E. Zakharov.
\newblock Five-wave interaction on the surface of deep fluid.
\newblock {\em Physica D: Nonlinear Phenomena}, 87(1-4):233--261, (1995).

\bibitem{Brazhnikov2002}
M.~Y. Brazhnikov, G.~V. Kolmakov, A.~A. Levchenko, and L.~P. Mezhov-Deglin.
\newblock Observation of capillary turbulence on the water surface in a wide
  range of frequencies.
\newblock {\em Europhys. Lett.}, 58(4):510--516, (2002).

\bibitem{Lukaschuk2009}
S.~Lukaschuk, S.~Nazarenko, S.~McLelland, and P.~Denissenko.
\newblock Gravity wave turbulence in wave tanks: space and time statistics.
\newblock {\em Physical review letters}, 103(4):044501, (2009).

\bibitem{Cobelli2011}
P.~Cobelli, A.~Przadka, P.~Petitjeans, G.~Lagubeau, V.~Pagneux, and A.~Maurel.
\newblock Different regimes for water wave turbulence.
\newblock {\em Physical Review Letters}, 107(21):214503, (2011).

\bibitem{Falcon2022}
E.~Falcon and N.~Mordant.
\newblock Experiments in surface gravity–capillary wave turbulence.
\newblock {\em Annual Review of Fluid Mechanics}, 54(1):1--25, (2022).

\bibitem{Dyachenko1992}
S~Dyachenko, A.C. Newell, A~Pushkarev, and V.E. Zakharov.
\newblock Optical turbulence: weak turbulence, condensates and collapsing
  filaments in the nonlinear schr{\"o}dinger equation.
\newblock {\em Physica D: Nonlinear Phenomena}, 57(1-2):96--160, (1992).

\bibitem{Bortolozzo2009}
U.~Bortolozzo, J.~Laurie, S.~Nazarenko, and S.~Residori.
\newblock Optical wave turbulence and the condensation of light.
\newblock {\em JOSA B}, 26(12):2280--2284, (2009).

\bibitem{Gurbatov2005}
S.N. Gurbatov, V.V. Kurin, L.M. Kustov, and N.V. Pronchatov-Rubtsov.
\newblock Physical modeling of nonlinear sound wave propagation in oceanic
  waveguides of variable depth.
\newblock {\em Acoustical Physics}, 51(2):152--159, (2005).

\bibitem{Ryutova2003}
M.~Ryutova and T.~Tarbell.
\newblock {MHD} shocks and the origin of the solar transition region.
\newblock {\em Physical review letters}, 90(19):191101, (2003).

\bibitem{David2022}
V.~David and S.~Galtier.
\newblock Wave turbulence in inertial electron magnetohydrodynamics.
\newblock {\em Journal of Plasma Physics}, 88(5):905880509, (2022).

\bibitem{Bisnovatyi1995}
G.S. Bisnovatyi-Kogan and S.A. Silich.
\newblock Shock-wave propagation in the nonuniform interstellar medium.
\newblock {\em Reviews of Modern Physics}, 67(3):661, (1995).

\bibitem{Devita2022}
F.~De~Vita, G.~Dematteis, R.~Mazzilli, D.~Proment, Y.~V. Lvov, and M.~Onorato.
\newblock Anomalous conduction in one-dimensional particle lattices:
  Wave-turbulence approach.
\newblock {\em Physical Review E}, 106(3):034110, (2022).

\bibitem{Zakharov1999}
V.~Zakharov.
\newblock Statistical theory of gravity and capillary waves on the surface of a
  finite-depth fluid.
\newblock {\em European Journal of Mechanics - B/Fluids}, 18(3):327--344,
  (1999).

\bibitem{Onorato2020}
M.~Onorato and G.~Dematteis.
\newblock A straightforward derivation of the four-wave kinetic equation in
  action-angle variables.
\newblock {\em Journal of Physics Communications}, 4(9):095016, (2020).

\bibitem{Yoshida1990}
H.~Yoshida.
\newblock Construction of higher order symplectic integrators.
\newblock {\em Physics Letters A}, 150(5-7):262--268, (1990).

\bibitem{Kramer2002}
P.~R. Kramer, J.~A. Biello, and Y.~Lvov.
\newblock Application of weak turbulence theory to {FPU} model.
\newblock {\em arXiv preprint nlin/0210007}, (2002).

\bibitem{Biello2002}
J.~A. Biello, P.~R. Kramer, and Y.~Lvov.
\newblock Stages of energy transfer in the {FPU} model.
\newblock {\em arXiv preprint nlin/0210008}, (2002).

\bibitem{Guasoni2017}
M.~Guasoni, J.~Garnier, B.~Rumpf, D.~Sugny, J.~Fatome, F.~Amrani, G.~Millot,
  and A.~Picozzi.
\newblock Incoherent {F}ermi-{P}asta-{U}lam recurrences and unconstrained
  thermalization mediated by strong phase correlations.
\newblock {\em Physical Review X}, 7(1):011025, (2017).

\bibitem{Toda1967}
M.~Toda.
\newblock Vibration of a chain with nonlinear interaction.
\newblock {\em Journal of the Physical Society of Japan}, 22(2):431--436,
  (1967).

\bibitem{Toda1975}
M.~Toda.
\newblock Studies of a non-linear lattice.
\newblock {\em Physics Reports}, 18(1):1 -- 123, (1975).

\bibitem{Goldfriend2019}
T.~Goldfriend and J.~Kurchan.
\newblock Equilibration of quasi-integrable systems.
\newblock {\em Phys. Rev. E}, 99(2), (2019).

\bibitem{Fu2019}
W.~Fu, Y.~Zhang, and H.~Zhao.
\newblock Universal law of thermalization for one-dimensional perturbed {T}oda
  lattices.
\newblock {\em New Journal of Physics}, 21(4):043009, Apr (2019).

\end{thebibliography}
\end{document}